\documentclass{acm_proc_article-sp}

\usepackage{paralist}
\usepackage{colortbl}
\usepackage{graphicx}
\usepackage{subfigure}
\graphicspath {{./} {Figures/}}
\newcolumntype{H}{>{\columncolor{black}\color{white}}c}
\newcommand{\col}[1]{\multicolumn{1}{H}{\bfseries #1}}
\begin{document}
\title{A Decade of Database Research Publications}

\numberofauthors{2}
\author{
\alignauthor Sherif Sakr\\
       \affaddr{National ICT Australia}\\
       \affaddr{University of New South Wales,  Australia}\\
       \email{ssakr@cse.unsw.edu.au}
\alignauthor Mohammad Alomari\\
       \affaddr{School of Information Technologies}\\
       \affaddr{University of Sydney,  Australia}\\
       \email{miomari@it.usyd.edu.au}
}

\date{30 July 2009}

\maketitle
\begin{abstract}
We analyze the database research publications of four major core
database technology conferences (SIGMOD, VLDB, ICDE, EDBT), two
main theoretical database conferences (PODS, ICDT) and three
database journals (TODS, VLDB Journal, TKDE) over a period of 10
years (2001 - 2010). Our analysis considers only regular papers as
we do not include short papers, demo papers, posters, tutorials or
panels into our statistics. We rank the research scholars
according to their number of publication in each
conference/journal separately and in combined. We also report
about the growth in the number of research publications and the
size of the research community in the last decade.

\end{abstract}

\section{Introduction}
\label{SecIntroduction} The database management technology has
played a vital role in the advancements of the information
technology field. Database researchers are one of the key players
and main sources to the growth of the database systems. They are
playing a foundational role in creating the technological
infrastructure from which database advancements evolve. The impact
of research scholars in the community is often measured by their
number of publications in top-tier research venues and the number
of citations they receive, i.e. how frequently their publications
are referenced by other publications (e.g. H-index~\cite{Hindex},
g-index~\cite{Gindex}). In principle, there is a direct
relationship between the tier rank of a research venue and its
number of citations which is commonly determined as the
\emph{impact factor}~\cite{ImpactFactor}. The success of a
research scholar in publishing  his research results in a top-tier
venue increases his chances of having his work being widely
received by his peers in the community and consequently to be more
frequently cited by them.

In general, achieving an accurate, fair and insightful
citation-based analysis is a very challenging task due to the
difficulty of parsing and extracting the citation meta data from
the research articles. Recently, some online services have been
introduced to capture the citation information of research
publications (e.g. MS
Libra\footnote{http://academic.research.microsoft.com/}, Google
Scholar\footnote{http://scholar.google.com.}). However, the
information provided by these services suffer from some anomalies
such as: incompleteness and duplication. Therefore, preparing a
high quality citation information for a pool of research
publications requires an extensive amount of manual labor work.
Moreover, citation-based analysis methods tend to consider only
the explicit citation relationships as indicated in the reference
parts of the articles. In practice, it is impossible for authors
of any article (including this one) to cite all the related
publications of their work but they  are normally only able to
cite only a fraction of them. Therefore, the final decision of
selecting the set of papers to be referenced usually depends on
many scientific and non-scientific factors. For example, it has
been shown that citations tend to have problems like
biased-citation, self-citation, or positive vs. negative
citation~\cite{CitationProblems,RankingMethods}. One common
situation is that article introductions are usually citing related
survey papers. Therefore, survey papers usually have citation
counts that are many times more than any original work in its
corresponding topic (e.g. according to Google Scholar, at the time
of writing this paper, the two surveys:~\cite{Survey1} has 883
citations and~\cite{Survey2} has 2169 citations). Some studies
have also shown that different citation choices correspond to
different citation impact~\cite{CitationImpact}.

Complementary to a previous work which mainly considered ranking
the research scholars based on their citation
counts~\cite{Rahm05}, in this paper, we focus on ranking the
research scholars by the count of their research publications in
top-tier venues. We selected a set of top-tier database research
venues which are generally considered as the most representative,
influential and prestigious in the database community. In
particular, we analyzed the database research publications of four
major core database technology conferences (SIGMOD, VLDB, ICDE,
EDBT), two main theoretical database conferences (PODS, ICDT) and
three database journals (TODS, VLDB Journal, IEEE TKDE) over a 10
years period (2001 - 2010). In general, we believe that research
fields are better presented by their own venues rather than by
multi-disciplinary venues. Therefore, we did not include some
important conferences (e.g. CIKM, WWW) and journals (e.g.
Information Systems) in the scope of this study.

In principle, some could argue that the number of publications may
have become a less insightful or less significant metric due to
the explosion of the number of conferences and journals in recent
years~\cite{Fabio}. Therefore, to remedy this argument, we
considered only top-tier venues which are well-known with their
very low \emph{acceptance rates}. These \emph{prestigious} venues
are conducting highly selective review processes that mainly aims
of ensuring that they are turning out high quality papers. Hence,
these papers are usually expected to attract considerable
attention (and citations) from other researchers in the
community~\cite{SelectivityImpact}. In fact, the distribution of
our selected venues (6 conferences and 3 journals) is compatible
with the fact that database researchers - and computer scientists
in general - are considering prestigious conferences as  favorite
tools for presenting original research work in contrast to the
general case of many other scientific disciplines where journal
papers are routinely considered to be superior than conference
papers~\cite{Growup,Growup2}. For example, it has been shown that
the two top database conferences (SIGMOD and VLDB) receive many
more citations per paper than the two top database journals (TODS
and VLDB J.)~\cite{Rahm05}. In practice, the general culture in
the computer science community is that journal papers are used to
present deeper versions of papers that already have been presented
at conferences. One of the main reasons behind this is that the
review process of journal papers are usually very long. The
\emph{turnaround time} (the interval between the submission date
of a manuscript and the date of having the editorial decision) for
conferences is often less than a third of that of
journals~\cite{JournalRelevance}. Since the field of computer
science research tends to be fast paced, conferences provide a
great chance for \emph{timestamping} the latest research findings
earlier which allows the knowledge to be publicly shared more
rapidly.

In general, we are witnessing a continuous growth in the database
field. That is mainly due to the continuous introduction of new
application domains (e.g. web applications, mobile applications,
cloud computing, sensor networks) with varying features and
requirements on their data management aspects. In practice, data
has become mobile, flexible, mirrored in a variety of logical and
physical forms, evolving, being concurrently modified and
replicated, dynamically generated and later reintegrated in very
large repositories for further analysis and
processing~\cite{DBMillennium}. Therefore, there are many more
researchers are entering the field to tackle these challenges and
hence more research papers are being published. In this paper, we
also study the growth rate on the size of contributing research
community and the number of research publications in the last
decade.

 The input data of this study has been extracted from
the XML records of the famous DBLP computer science
bibliography\footnote{http://dblp.uni-trier.de/xml/}. Our analysis
considers only regular papers as we do not include short papers,
demo papers, posters, tutorials or panels into our statistics. We
made the detailed results of our study accessible on the
web\footnote{http://www.cse.unsw.edu.au/$\sim$ssakr/DBStatistics/index.html}

\vspace{-0.25cm}
\section{Study Results}
\subsection{Top Publishers of Database Research Venues} \label{SecTop}

\begin{figure*}
\centering \subfigure[VLDB] {
    \label{FigVLDB10}
    \includegraphics[width=0.45\textwidth,height=2.5in]{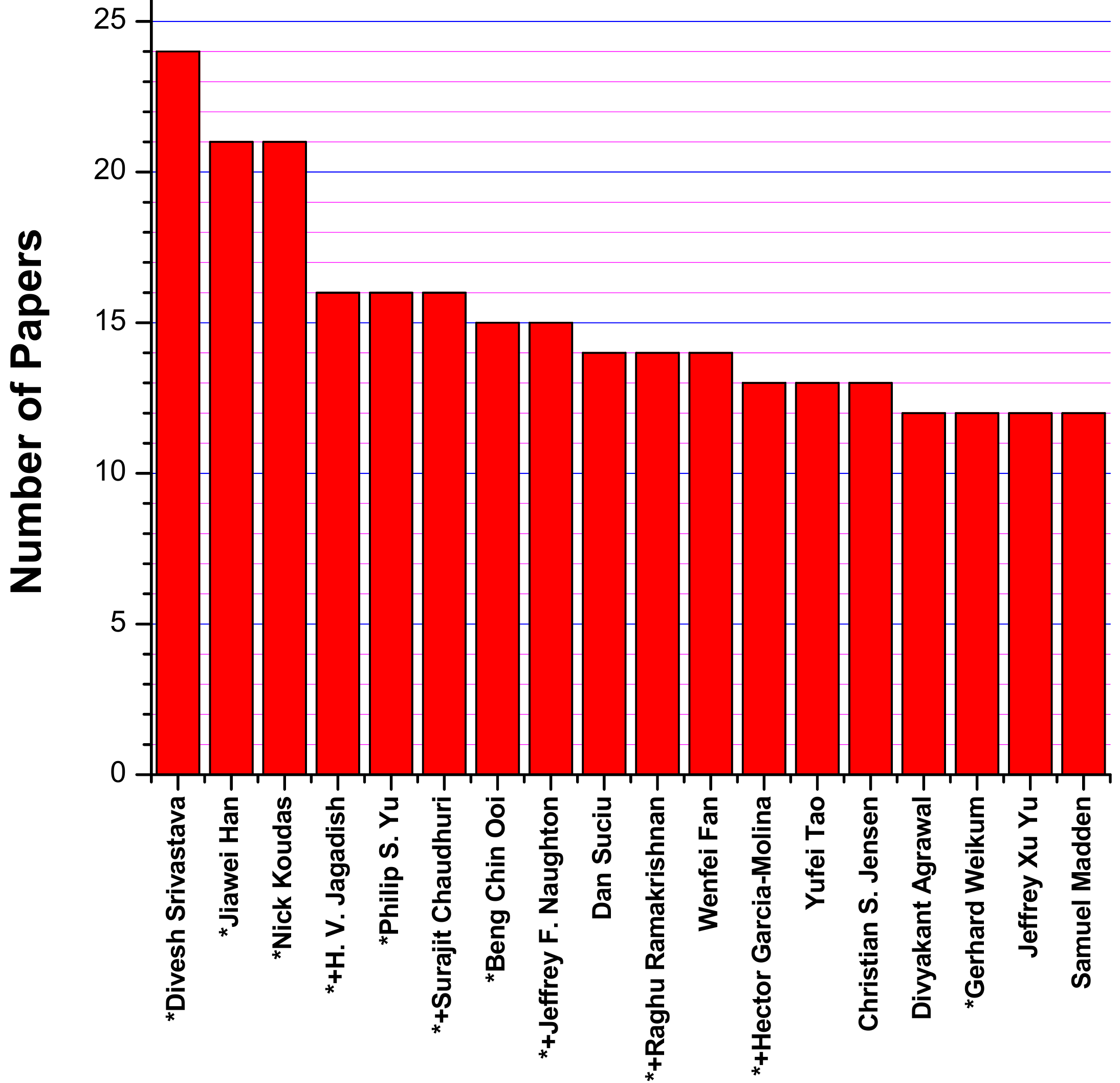}
} \subfigure[SIGMOD] {
    \label{FigSigmod10}
      \includegraphics[width=0.45\textwidth,height=2.5in]{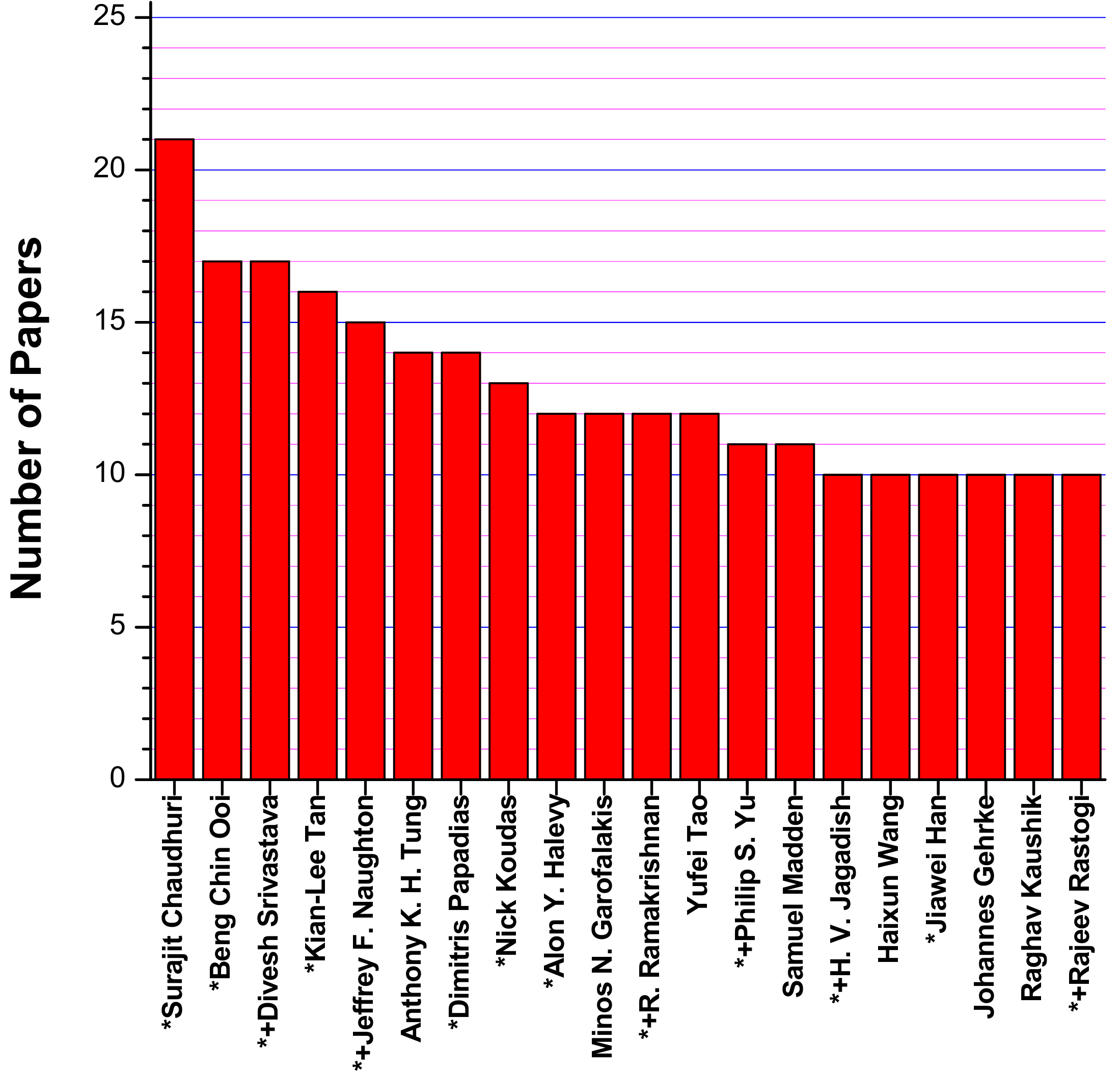}
} \subfigure[ICDE] {
    \label{FigICDE10}
      \includegraphics[width=0.45\textwidth,height=2.5in]{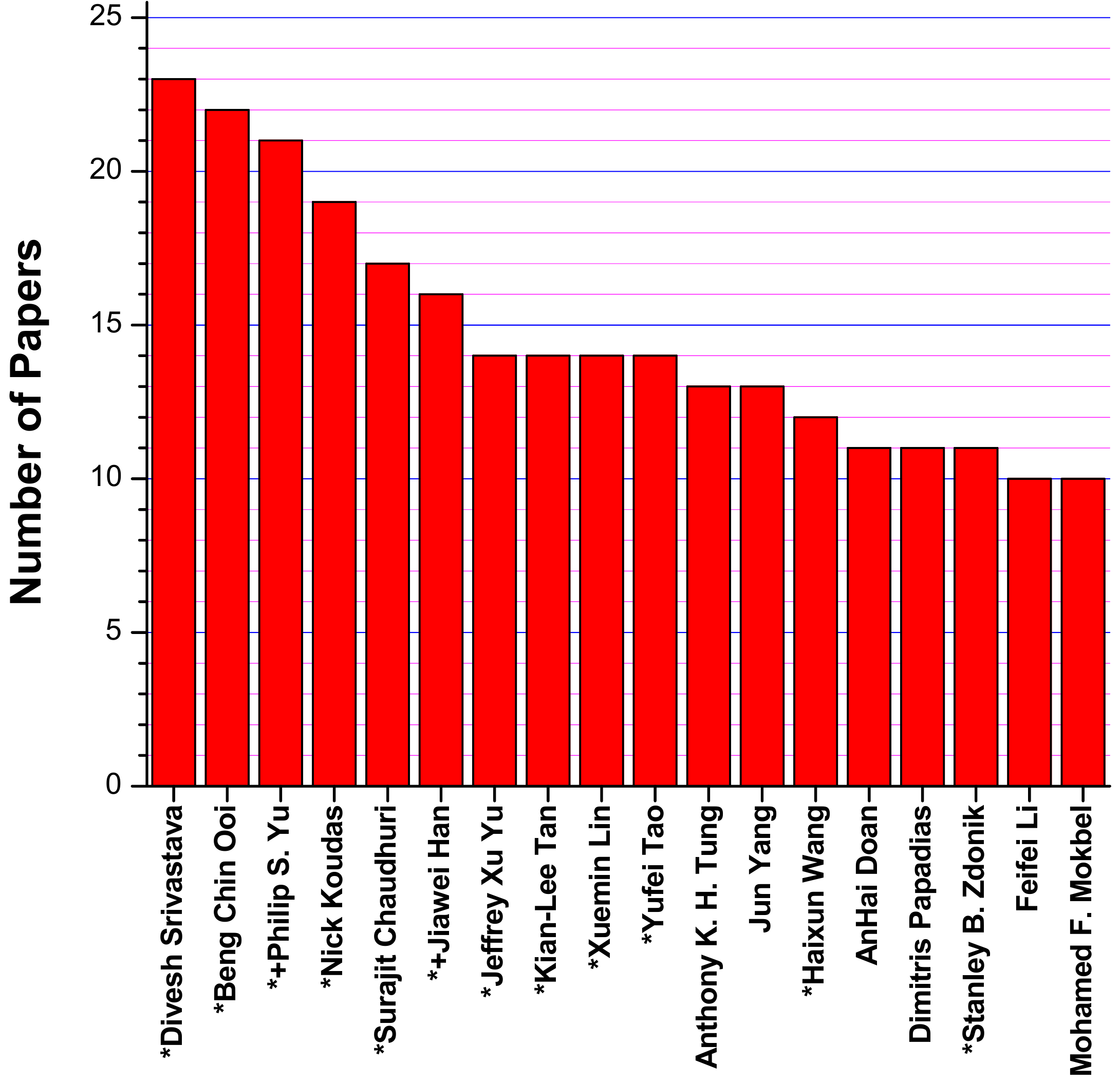}
}\subfigure[EDBT] {
    \label{FigEDBT10}
      \includegraphics[width=0.45\textwidth,height=2.5in]{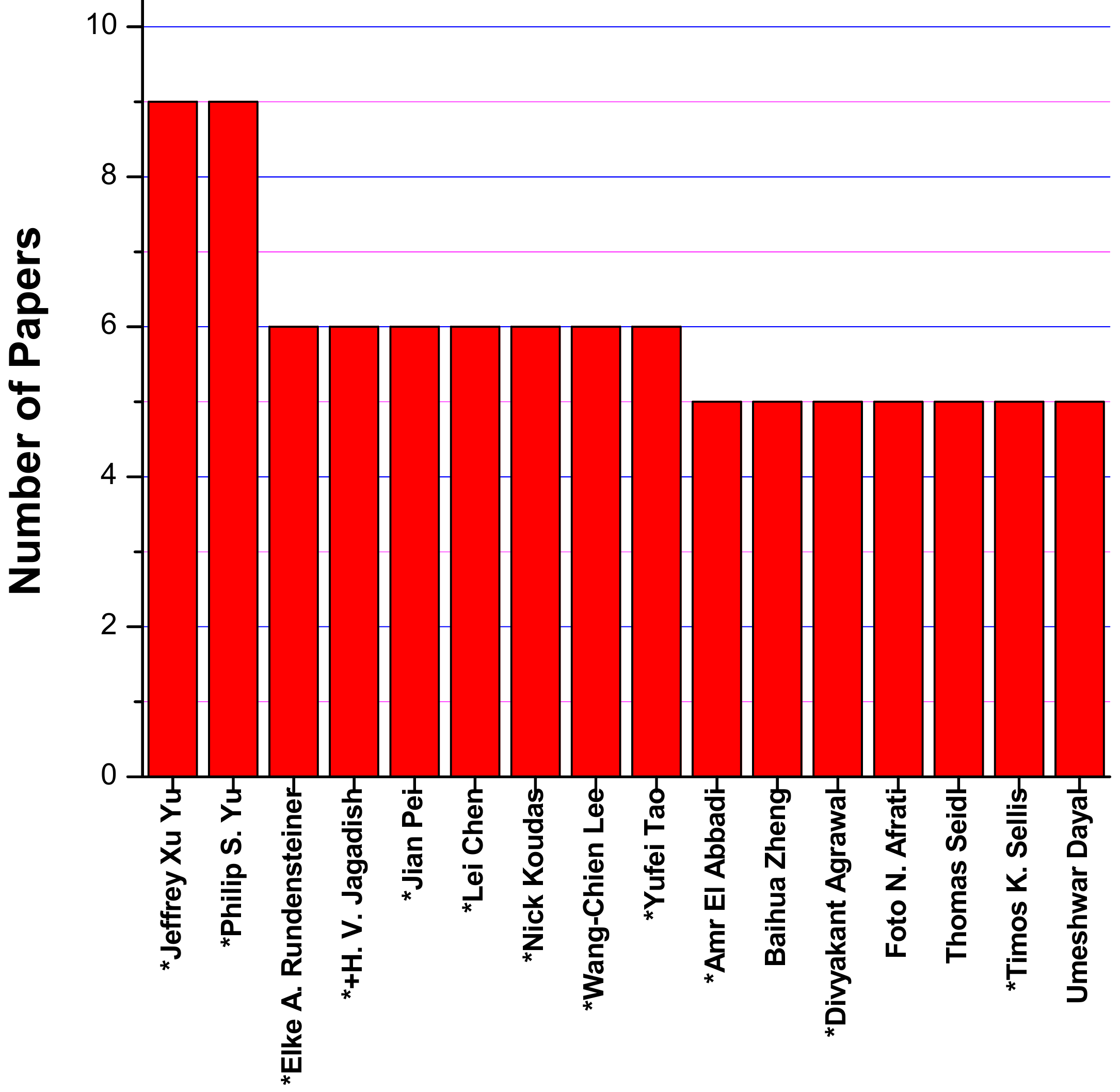}
}\caption{Top Publishers in Major Core Database Technology
Conferences} \label{FigMajorDB}
\end{figure*}

\begin{figure*}
\centering \subfigure[PODS] {
    \label{FigPODS10}
    \includegraphics[width=0.45\textwidth,height=2.5in]{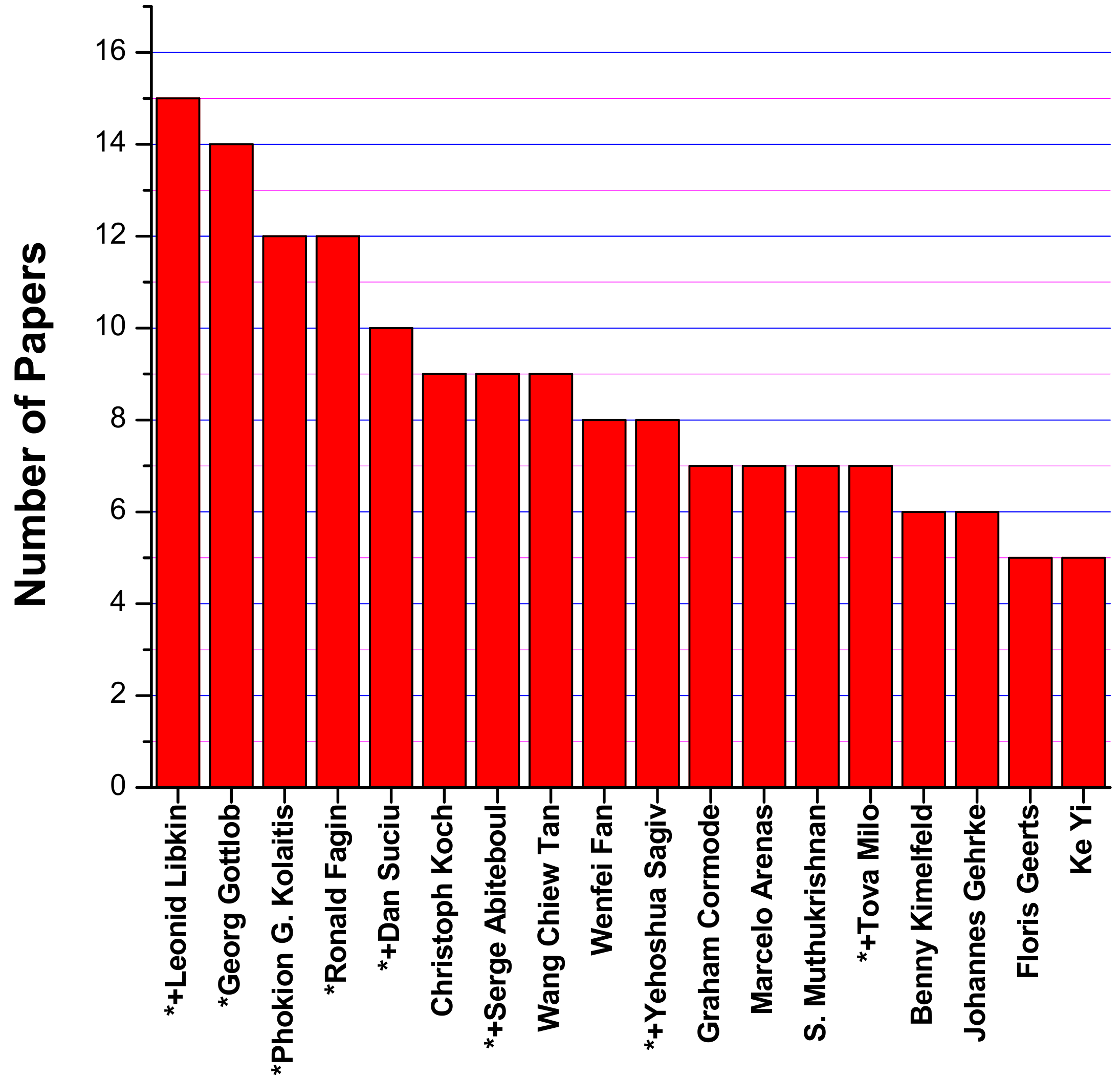}
} \subfigure[ICDT] {
    \label{FigICDT10}
      \includegraphics[width=0.45\textwidth,height=2.5in]{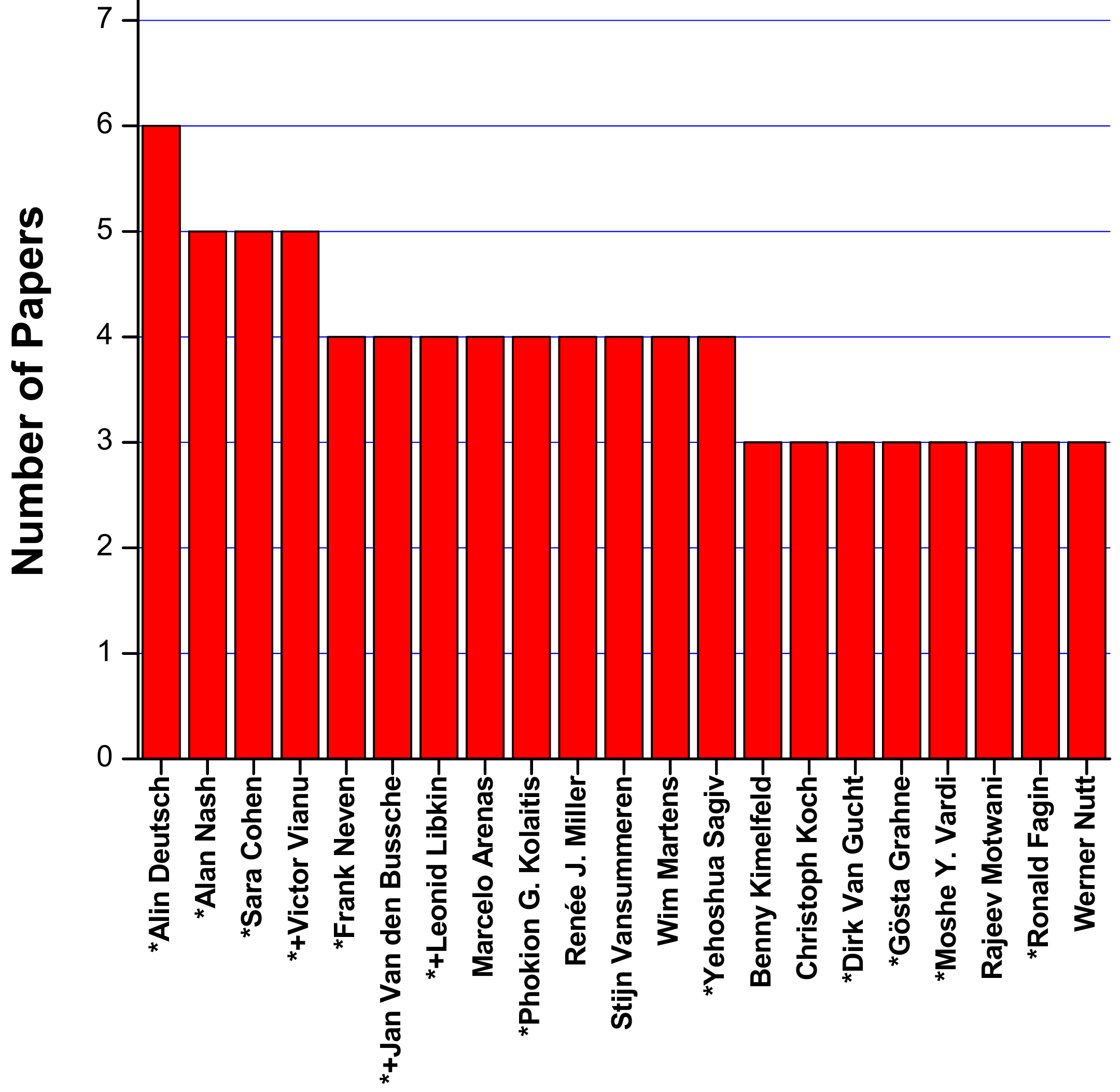}
} \caption{Top Publishers in Major Theoretical Database
Conferences} \label{FigTheorticalDB}
\end{figure*}

\begin{figure*}
\centering \subfigure[VLDB J.] {
    \label{FigVLDBJ10}
    \includegraphics[width=0.31\textwidth,height=2.5in]{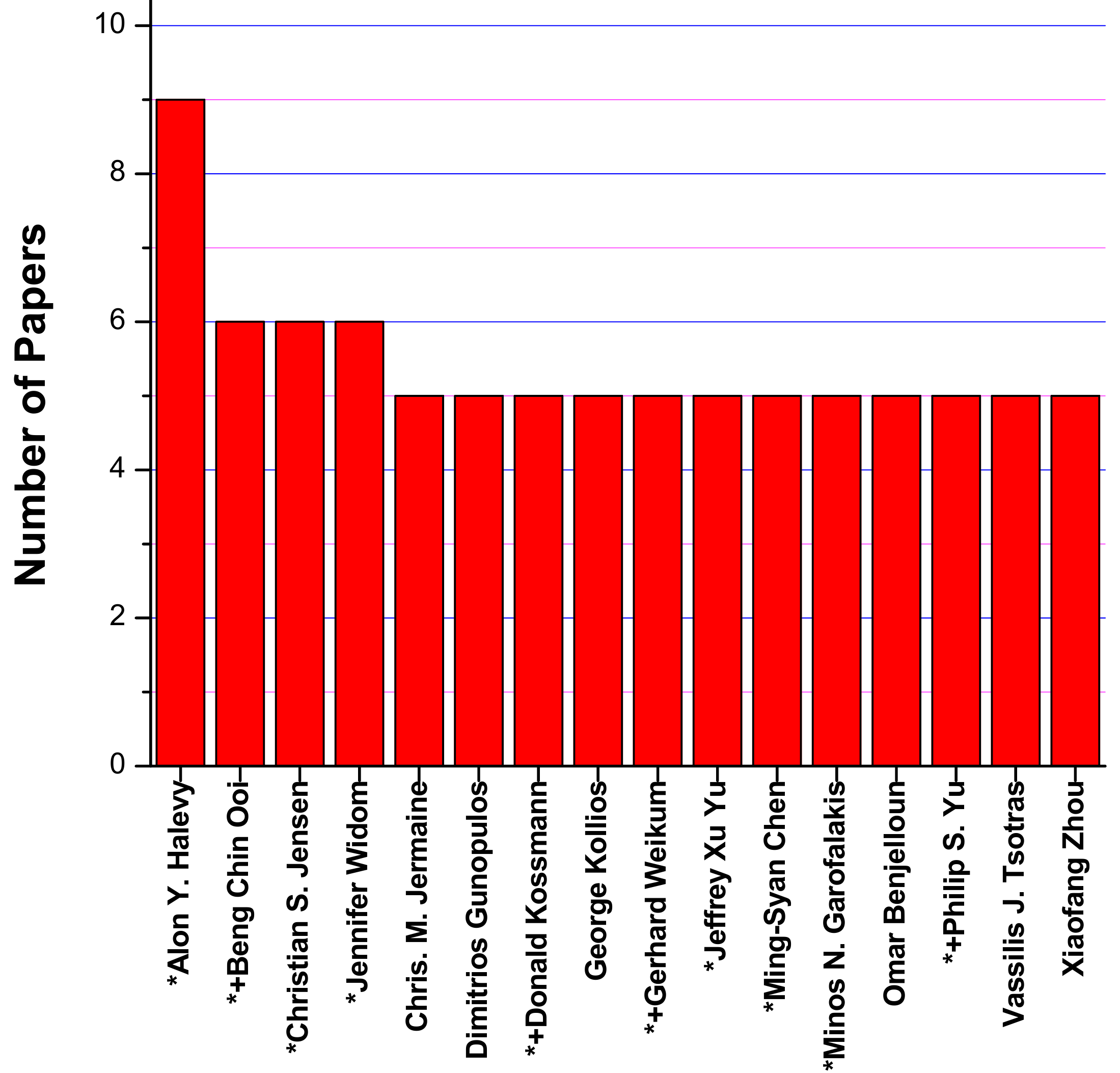}
} \subfigure[TODS] {
    \label{FigTODS10}
      \includegraphics[width=0.31\textwidth,height=2.5in]{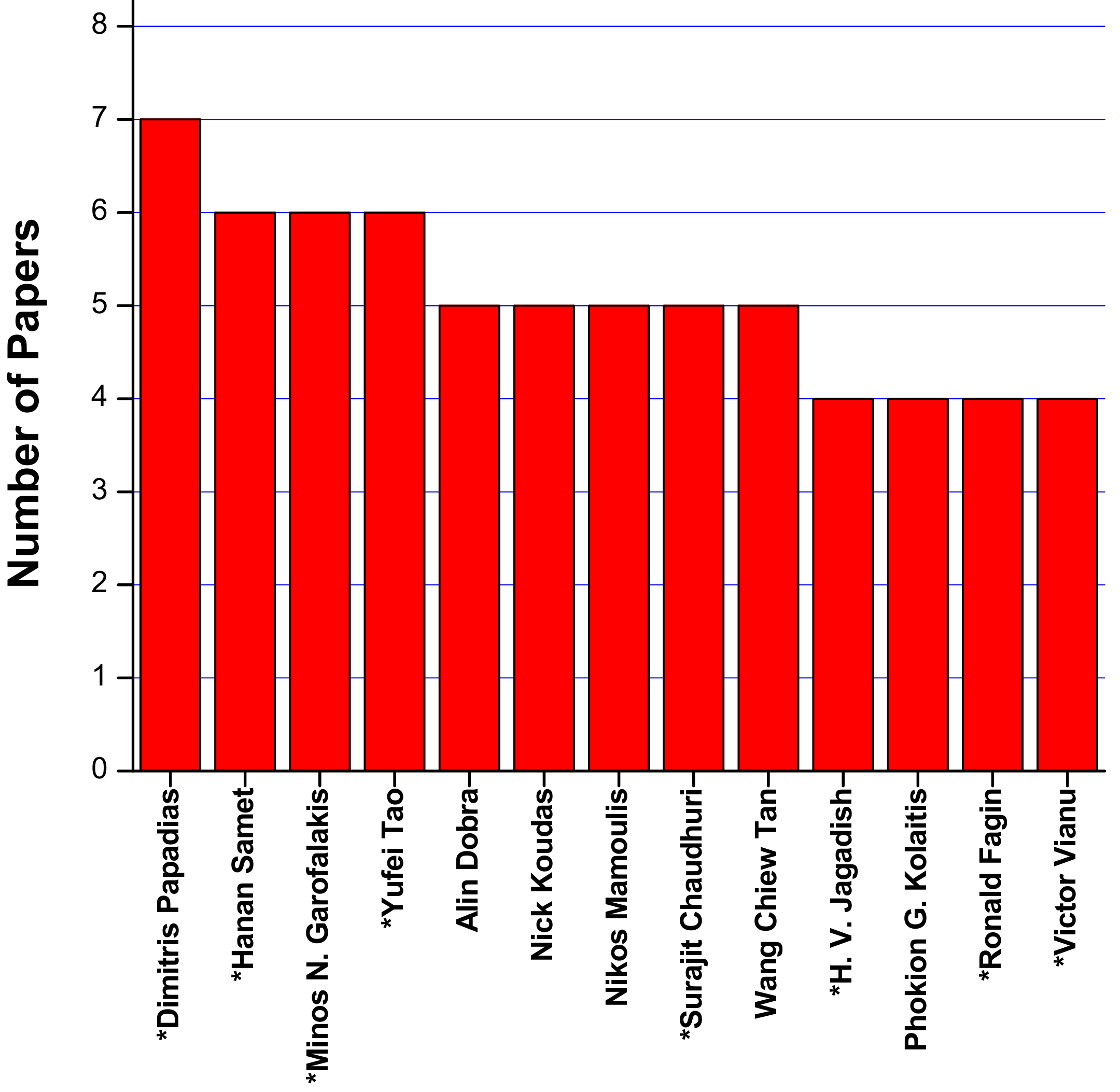}
}\subfigure[TKDE] {
    \label{FigTKDE10}
      \includegraphics[width=0.31\textwidth,height=2.5in]{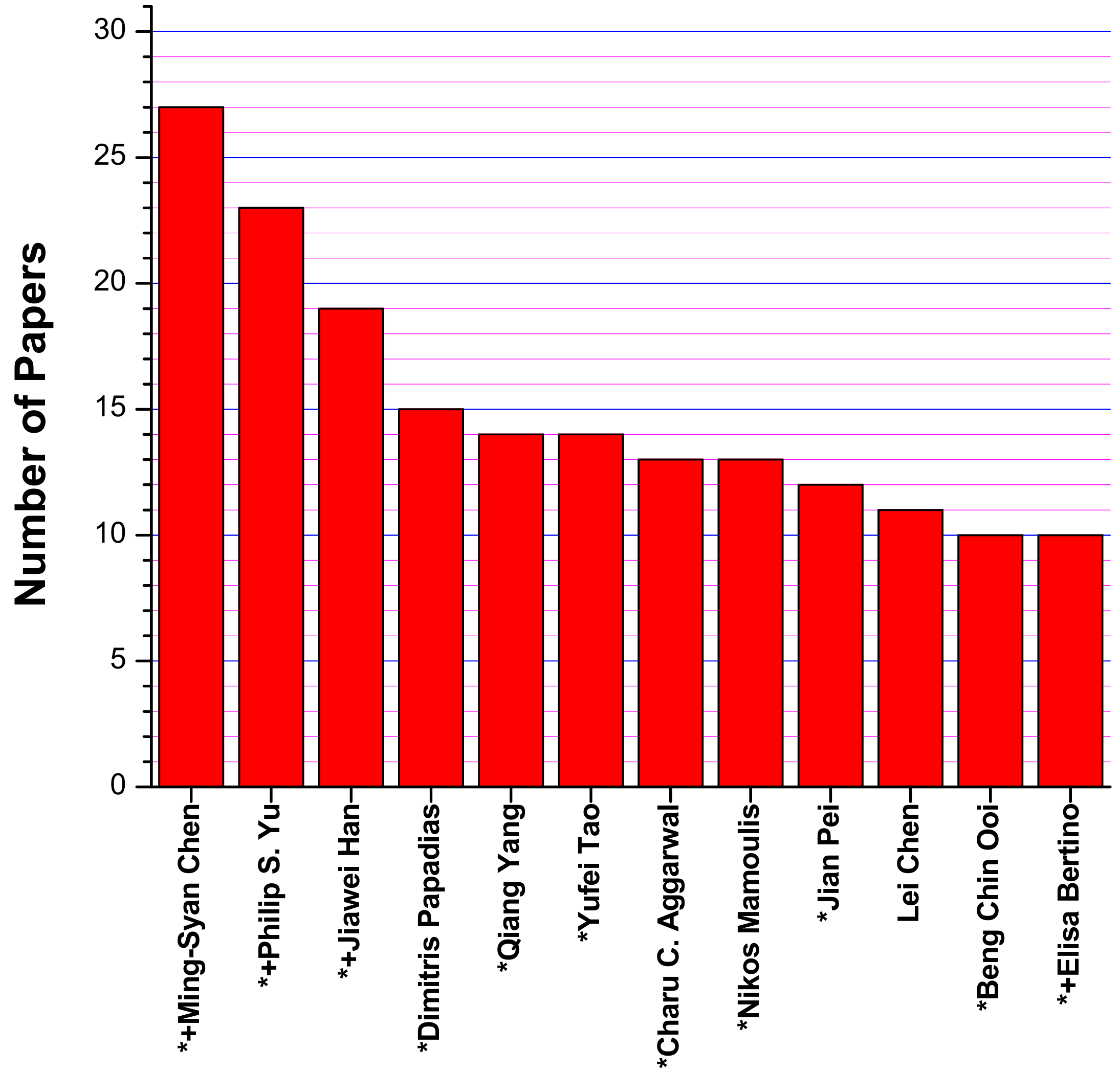}
} \caption{Top Publishers in Major Database Technology Journals}
\label{FigDBJournal}
\end{figure*}

\begin{figure*}
\subfigure[Core DB: VLDB + SIGMOD + ICDE + EDBT] {
    \label{FigCombinedCore10}
    \includegraphics[width=0.31\textwidth,height=2.5in]{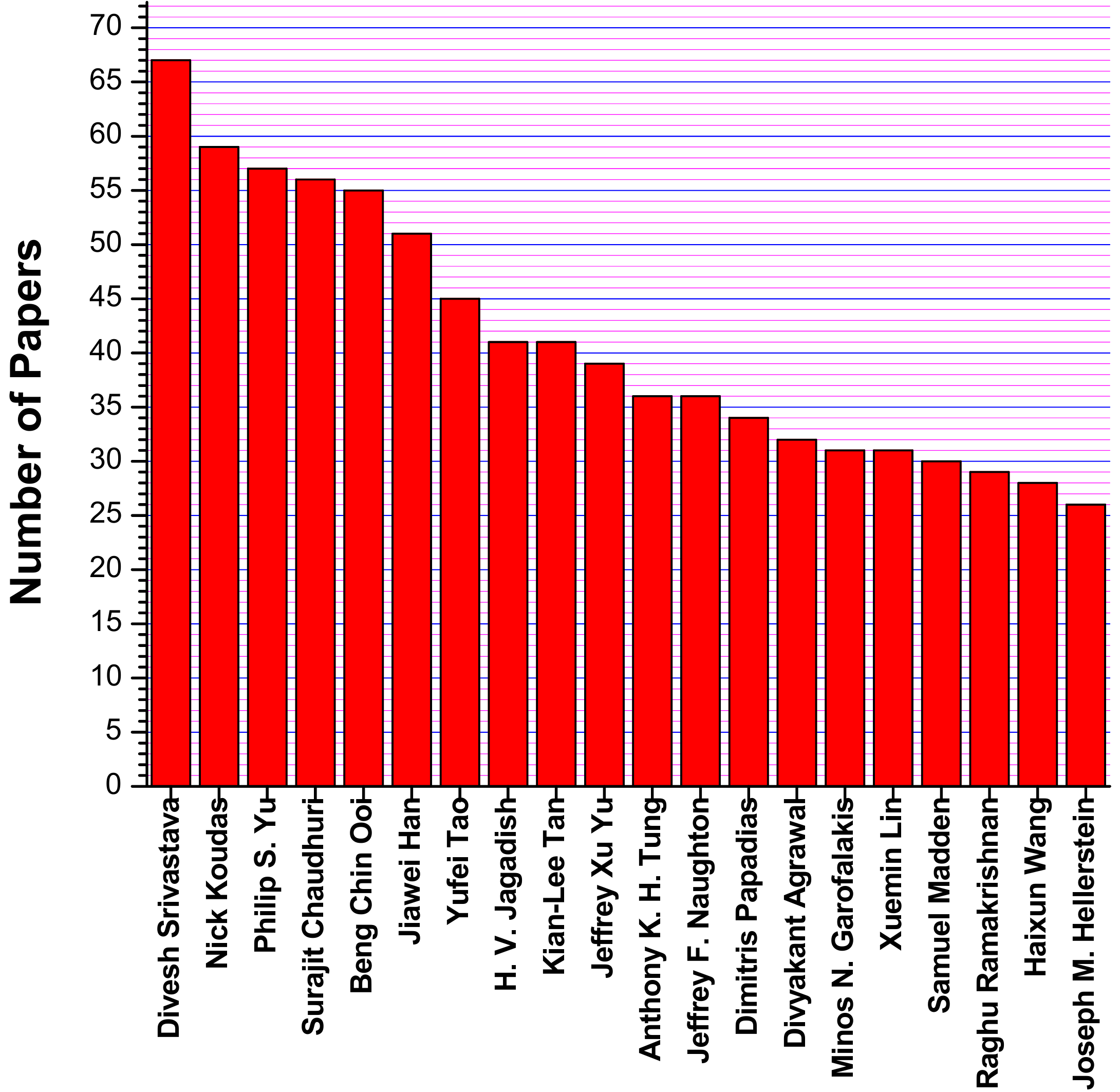}
} \subfigure[Theoretical DB: PODS + ICDT] {
    \label{FigCombinedTheory10}
      \includegraphics[width=0.31\textwidth,height=2.5in]{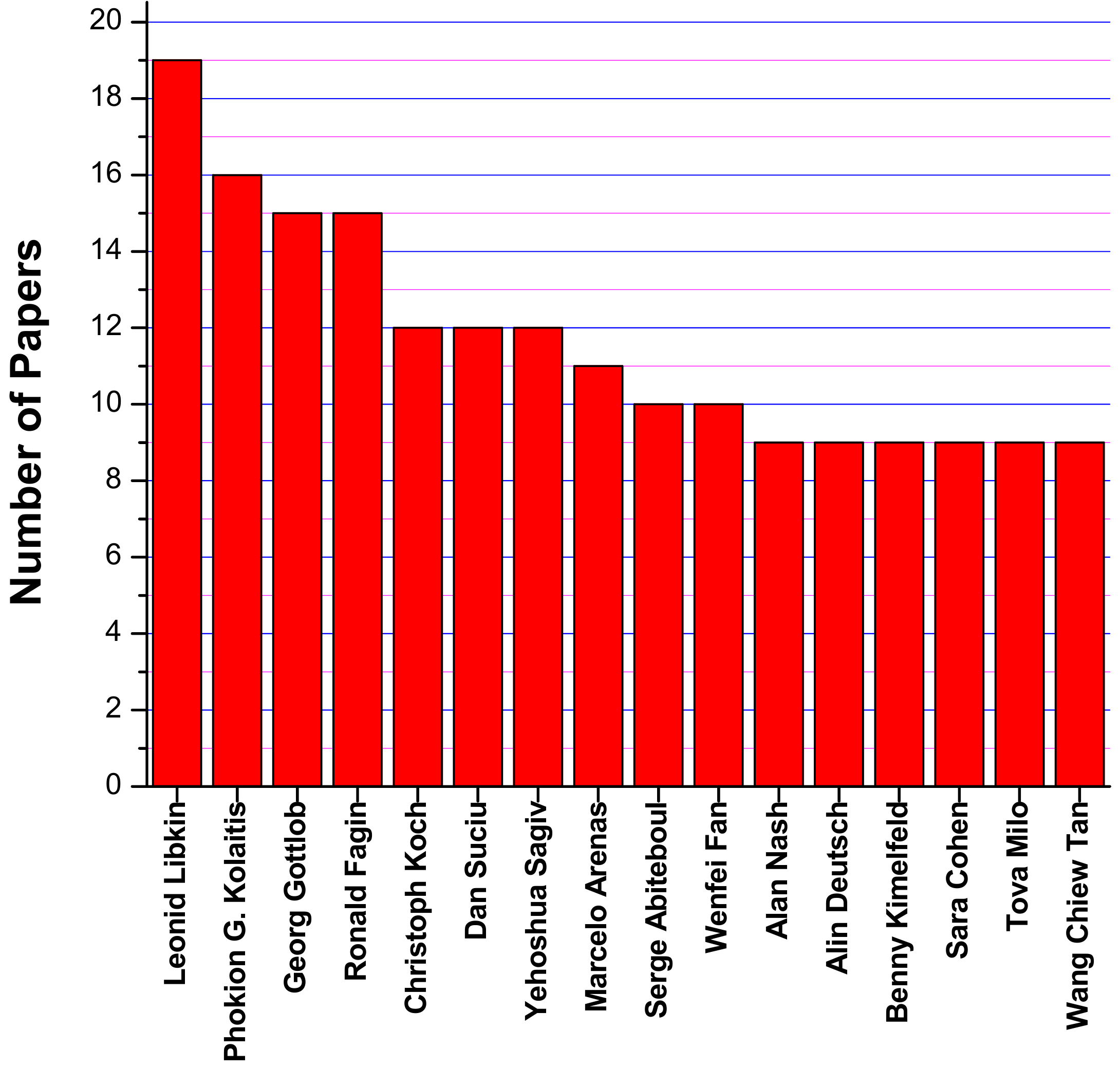}
} \subfigure[DB Journals: VLDB J. + TODS + TKDE] {
    \label{FigCombinedJournal10}
      \includegraphics[width=0.32\textwidth,height=2.5in]{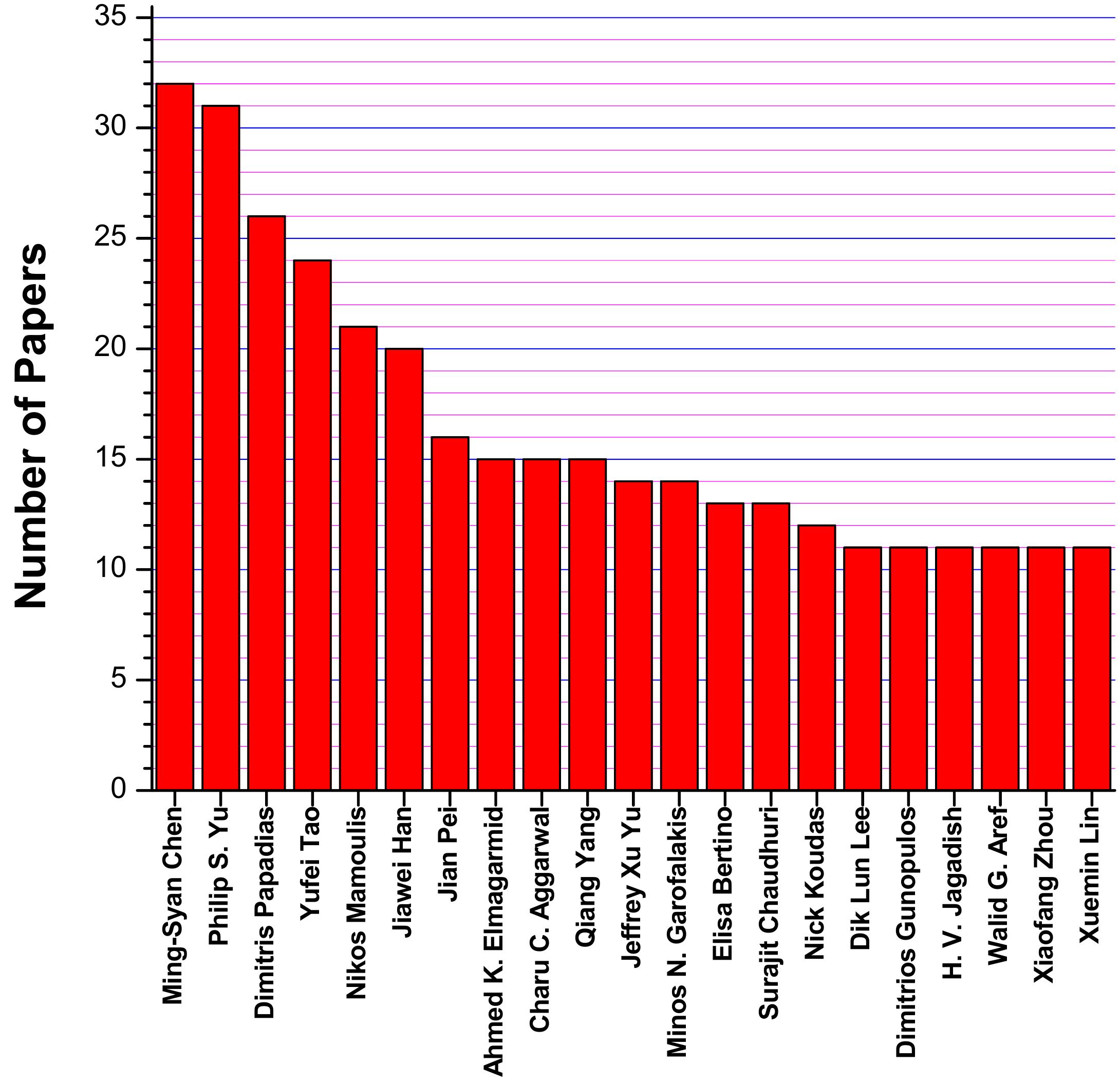}
} \caption{Aggregate Lists of Top Publishers for Database
Research Venues} \label{FigCombined}
\end{figure*}
\begin{figure*}
\centering \subfigure[Conferences] {
    \label{Figsub}
    \includegraphics[width=0.47\textwidth,height=2.25in]{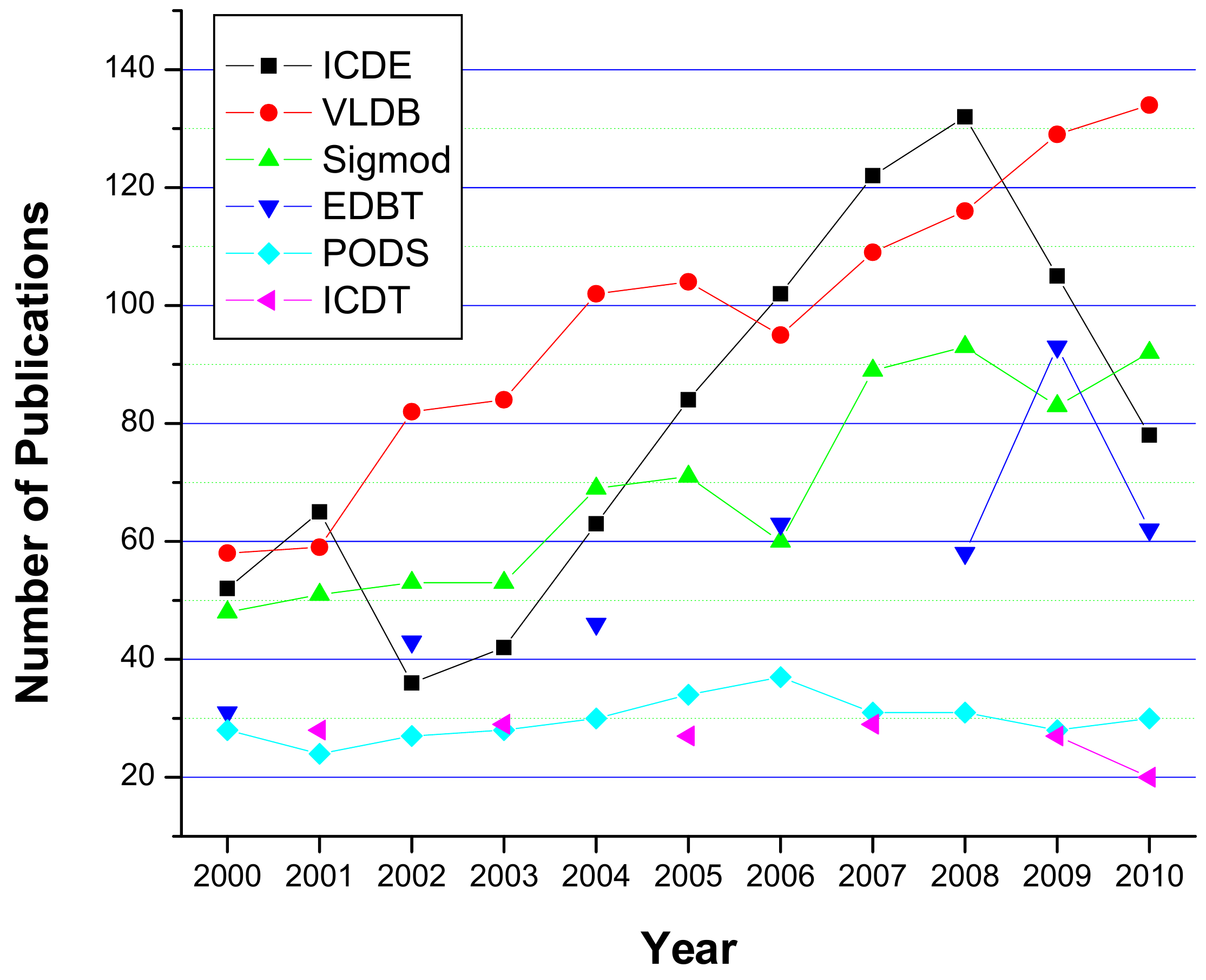}
}
\subfigure[Journals] {
    \label{FigSuper}
      \includegraphics[width=0.47\textwidth,height=2.25in]{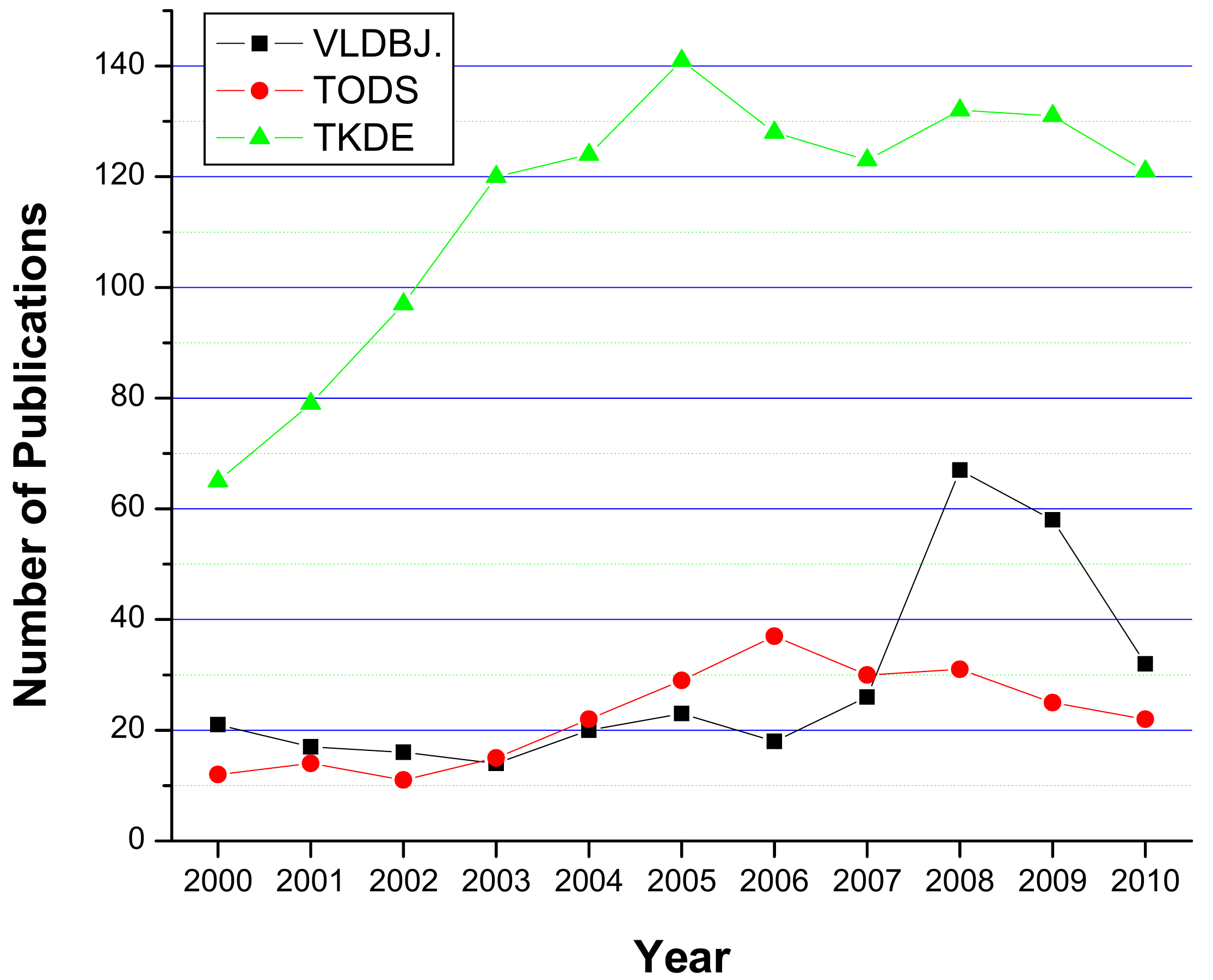}
} \caption{Growth in Number of Publications}
\label{FigGrowthPublications}
\end{figure*}

\begin{figure*}
\centering \subfigure[Conferences] {
    \label{Figsub}
    \includegraphics[width=0.47\textwidth,height=2.25in]{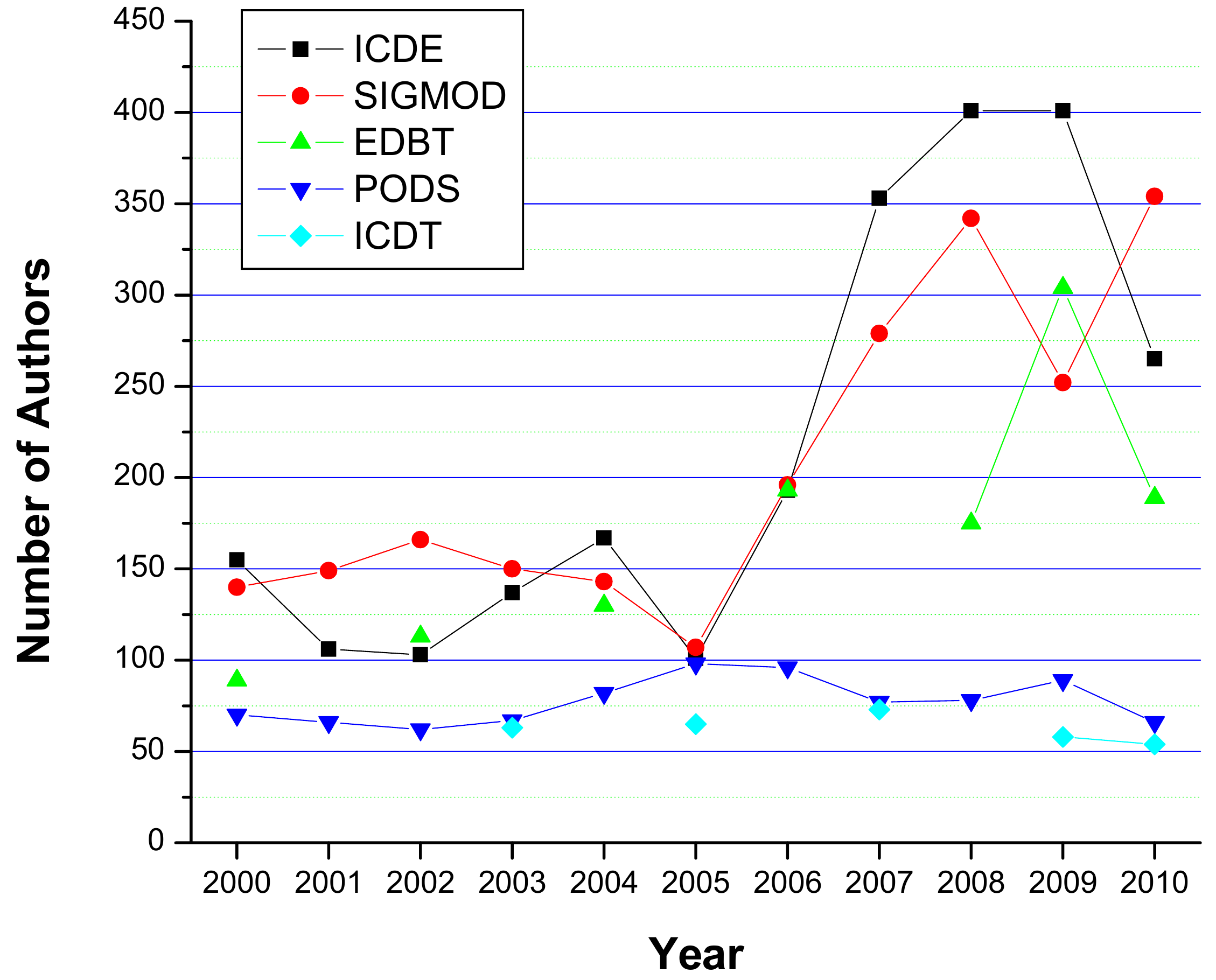}
}
\subfigure[Journals] {
    \label{FigSuper}
      \includegraphics[width=0.47\textwidth,height=2.25in]{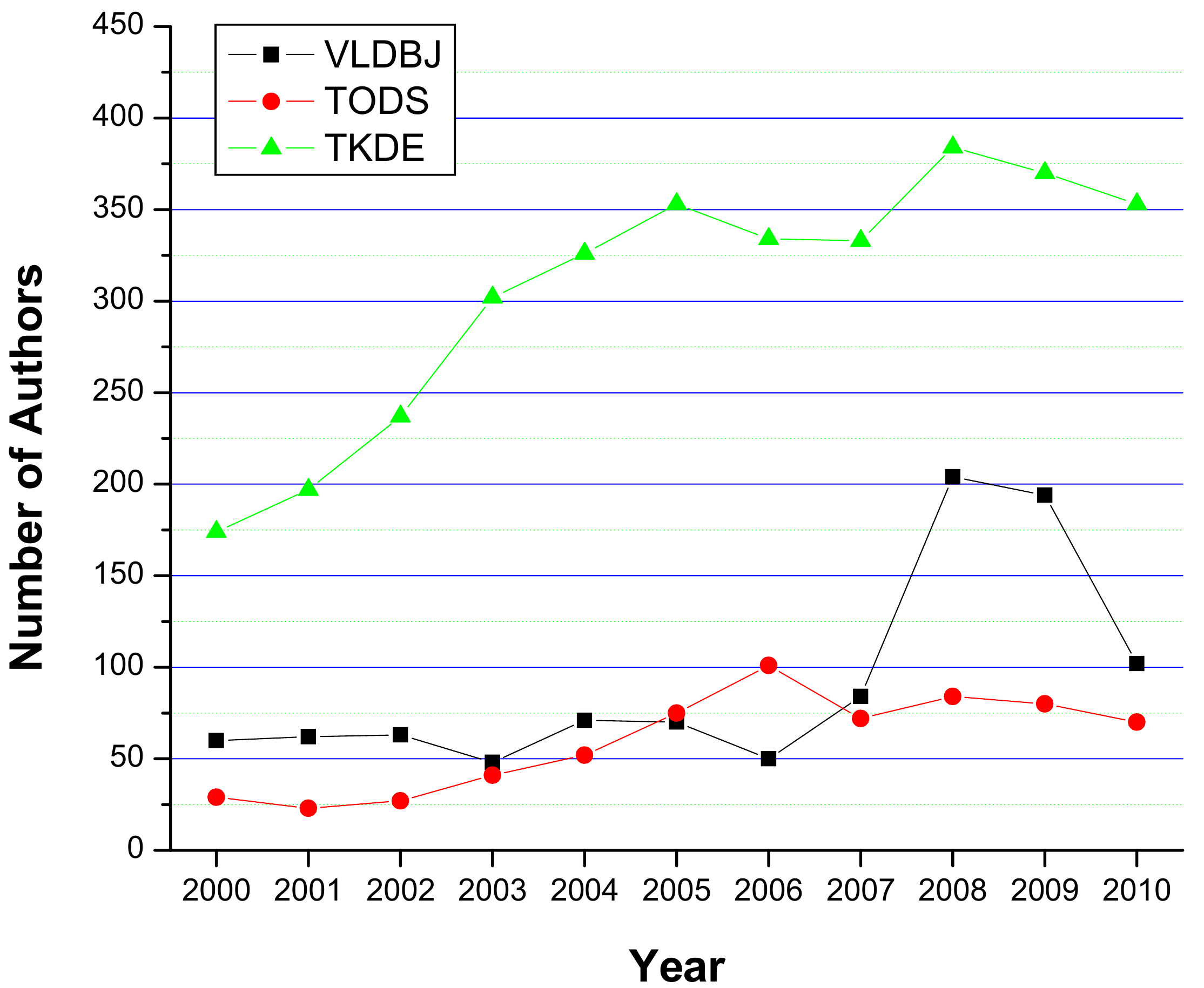}
} \caption{Growth in Number of Authors} \label{FigGrowthAuthors}
\end{figure*}
As we previously stated, in this study, we focus on measuring the
number of publications in top-tier publication venues as one of
the main indicators to evaluate the impact of a research scholar
in the community and the quality of his research production. In
this paper, we present the most important results of our study.
For full detailed results, we refer the reader to the web page of
this study.

Figures~\ref{FigMajorDB},~\ref{FigTheorticalDB},~\ref{FigDBJournal}
illustrate the top publishers of the database research venues
during the period between 2001 and 2010. Figure~\ref{FigMajorDB}
represents the top publishers of the core database technology
conferences: VLDB (Figure~\ref{FigVLDB10}), SIGMOD
(Figure~\ref{FigSigmod10}), ICDE (Figure~\ref{FigICDE10}) and EDBT
(Figure~\ref{FigEDBT10}). Figure~\ref{FigTheorticalDB} represents
the top publishers of the theoretical database conferences: PODS
(Figure~\ref{FigPODS10}) and ICDT (Figure~\ref{FigICDT10}).
Figure~\ref{FigDBJournal} represents the top publishers of the
main database journals: VLDB journal (Figure~\ref{FigVLDBJ10}),
TODS journal (Figure~\ref{FigTODS10}) and TKDE
(Figure~\ref{FigTKDE10}). The research scholars in these figures
can be indicated with one of the following two symbols:
\begin{compactitem}
    \item The (+) symbol indicates that the research scholar
    appears on the correspondingly top publishers list of the same
    research venue for the former decade (1991 - 2000).
    \item The (*) symbol indicates that the research scholar
    appears on the \emph{ultimate} top publishers list of the same
    research venue in all of its editions since its origin.
\end{compactitem}
For example, in Figure~\ref{FigVLDB10}, \emph{Divesh Srivastava}
and  \emph{H. V. Jagadish} are indicated that they appear in the
top publishers of the VLDB conference since its origin (1975 -
2010). However, only \emph{H. V. Jagadish} is indicated that he
appears on top publishers list of the VLDB conference on the
former decade. Figure~\ref{FigCombined} illustrates aggregate
lists of the top publishers for database research venues according
to their focus: core database technology conference
(Figure~\ref{FigCombinedCore10}), theoretical database conferences
(Figure~\ref{FigCombinedTheory10}) and database journals
(Figure~\ref{FigCombinedJournal10}). Several remarks can be
observed from the reported results for these database research
venues. Some key remarks are given as follows:

\begin{compactitem}

    \item There are distinctly 42 (non-distinctly 72) research scholars in the top publishers
    lists of the four core database technology conferences. There are distinctly
    34 (non-distinctly 41) research scholars in the top publishers
    lists of the three main database journals. In combination,
    there are 63 distinct research scholars on the seven venues.
    These results show a clear overlap between the list of these
    top database research venues.

    \item Three research scholars appear on the top publishers list of
    \emph{all} core database technology conferences. Namely, \emph{Philip S.
    Yu}, \emph{Nick Koudas} and \emph{Yufei Tao}. In addition, \emph{Philip S.
    Yu} appears on the top publishers lists of the VLDB journal and
    TKDE. \emph{Yufei Tao} appears on the lists of the TODS
    and TKDE while \emph{Nick Koudas} appears only on the
    list of TODS.

    \item Six research scholars appear on the top publishers list of
    \emph{three} (out of four) core database technology conferences. Namely, \emph{Divesh Srivastava}, \emph{Beng Chin Ooi}, \emph{Surajit Chaudhuri}, \emph{Jiawei Han}, \emph{Jeffrey Xu Yu} and \emph{H. V.
    Jagadish}. In addition, \emph{Beng Chin Ooi} appears on the
    lists of the VLDB Journal and TKDE.  \emph{Jiawei Han} appears on the top list of
    TKDE. \emph{Jeffrey Xu Yu} appears on the top list of the
    VLDB Journal. \emph{Surajit Chaudhuri}
    and \emph{H. V. Jagadish} appears on the top list of TODS.

    \item Eight research scholars  appear on the top publishers list of
    \emph{two} core database technology conferences. Namely, \emph{Kian-Lee
    Tan}, \emph{Anthony K. H. Tung}, \emph{Haixun Wang}, \emph{Dimitris Papadias}, J\emph{effrey F.
    Naughton},\emph{ Raghu Ramakrishnan}, \emph{Divyakant Agrawal} and \emph{Samuel
    Madden}. In addition, \emph{Dimitris Papadias} appears on the
    top publishers lists of TODS and TKDE.

    \item There are 32 distinct research scholars in the top
    publishers list of the two theoretical database conferences (PODS and
    ICDT). Seven research scholars appear on the lists of both
    conferences. Namely, \emph{Leonid Libkin}, \emph{Marcelo Arenas}, \emph{Phokion G.
    Kolaitis}, \emph{Yehoshua Sagiv}, \emph{Benny Kimelfeld}, \emph{Christoph Koch}
    and \emph{Ronald Fagin}.

    \item Seven research scholars have joint appearance on the top
    publishers
    list of at least one of the theoretical database conferences
    in addition to
    another appearance in at least one the top publishers list of a core database technology
    conference or a main database journal. Namely, \emph{Victor Vianu} (ICDT,
    TODS), \emph{Phokion G. Kolaitis} (PODS / ICDT, TODS), \emph{Ronald Fagin} (PODS / ICDT, TODS), \emph{Johannes Gehrke} (PODS, SIGMOD), \emph{Wang Chiew Tan} (PODS, TODS), \emph{Dan
    Suciu} (PODS, VLDB) and \emph{Wenfei Fan} (PODS, VLDB).

    \item \emph{Ming-Syan Chen} has the highest total number of
     publications in the major database journals in one year. In 2008, he has published 9 papers
    (5 papers in TKDE and 4 papers in VLDB Journal).

    \item \emph{Philip S. Yu} has the highest
    total number of  publications in the major database conferences in one year. In 2009, he has published 13 papers (6 papers in VLDB, 5 papers in ICDE and 2 papers in SIGMOD).

    \item \emph{Divesh Srivastava} is the top publisher in the
    aggregate list of all core database technology conferences
    (Figure~\ref{FigCombinedCore10}). He published 67 papers in
    total with an average of about 7 papers per year. On the other
    side, he published only 5 papers in the main database
    journals. Therefore, he does not appear in the aggregate list
    of the main database journals
    (Figure~\ref{FigCombinedJournal10}). Ten research scholars
    appear in both of the aggregate lists for top publishers on core database
    technology conferences and database journals. Namely, \emph{Philip S. Yu} (with total of 88 papers), \emph{Nick Koudas}
    (71 papers), \emph{Jiawei Han} (71 papers), \emph{Surajit Chaudhuri} (69 papers), \emph{Yufei
    Tao} (69 papers), \emph{H. V. Jagadish} (62 papers), \emph{Dimitris Papadias}
 (60 papers), \emph{Jeffrey Xu Yu} (53 papers), \emph{Minos N. Garofalakis} (45
 papers) and \emph{Xuemin Lin} (42 papers).

\item \emph{Yannis Papakonstantinou} and \emph{Dan Suciu} had at
least one paper in each of the studied nine major database venues
in the last decade.

\item Table 1 shows the most important co-authorship relations
between research scholars in the top lists of the database
research venues. For example, \emph{Yufei Tao }and \emph{Dimitris
Papadias} have participated in the co-authorship of 34 regular
paper in the different database research venues. The degree column
(Deg.) indicates the number of the research scholars participating
in the relationship.

\end{compactitem}

\begin{table}
\centering \scriptsize
\begin{tabular}{|l|l|c|}
  \hline
  \col{Deg.} &  \col{Authors} & \col{\# Pub.}\\
  \hline
  2 & Yufei Tao and Dimitris Papadias & 34\\\hline
    2 & Divesh Srivastava and  Nick Koudas & 33\\\hline
      2 & Divyakant Agrawal  and  Amr El Abbadi & 30\\\hline
        2 & Vivek R. Narasayya  and Surajit Chaudhuri & 22\\\hline
          2 & Beng Chin Ooi and  Anthony K. H. Tung & 16\\\hline
            2 & Haixun Wang and Philip S. Yu  & 16\\\hline
              2 & Xuemin Lin  and Wei Wang  &16\\\hline
              2 & Xuemin Lin  and Jeffrey Xu Yu &14\\\hline
  3 & B. Gedik, P. S. Yu and K. Wu & 9\\\hline
  3 & D. Agrawal, A. El Abbadi and A. Metwally & 7\\\hline
\end{tabular}
\caption{Top Co-authorship Relationships} \label{TBLSummary}
\end{table}

\vspace{-0.25cm}
\subsection{The Growth in number of Publications and Database
Community Size} \label{SecGrowth}

The topics of the database field is continuously growing.
Therefore, there are more researchers who are entering the
research community and more research papers are being
published~\cite{CommunityGrowth}. In our study, we determined the
number of regular publications for all of our considered
publication venues for the ten years period of 2001 - 2010.
Moreover, we determined the number of unique authors for the
publications of each venue as a measure of its contributing
community size. Figure~\ref{FigGrowthPublications} presents an
overview of the growth in the number of publications in the
database research venues while Figure~\ref{FigGrowthAuthors}
presents an overview of the growth in the number of unique authors
(participating community size). Combining the results of both
figures show that the number of research publications and unique
authors in core database technology conferences and database
journals has on average nearly doubled in number. On the contrary
for the theoretical database conference (PODS and ICDT), there was
no clear increase either on the number of publications nor on the
number of authors. They kept having an average of around 30 papers
and 75 authors per conference over the whole decade.

In principle, the number of regular research publications for core
database technology conferences cannot continue growing in
proportion to the size of the community. Therefore, most of the
conference have introduced other forms of publications such as:
posters, short papers and demo papers in order to provide a chance
for a wider part of the community to present their work and to
continue attracting and focusing the researchers to participate in
a small set of top conferences as there are always limits on the
number of conferences that researchers can attend. For example,
the 2002 edition of the ICDE conference first introduced the
acceptance of demo papers, the 2003 edition introduced the
acceptance of poster papers and the 2009 edition introduced the
acceptance of 4 pages short papers. We believe that having more
journal papers could be a good solution to absorb this continuous
increase of research publications without the need to increase the
number of conferences or to increase the number of accepted papers
in the current conferences.

One of the main reasons behind the increase in the number of
publications in the database community is the continuous
introduction of new research challenges which is relevant to the
scope of the community. For example, XML has started to be
introduced as a hot research topic for the database research
community in the early of the last decade. Moro et
al~\cite{XMLHaystack} referenced a list of more than 100
publications in a survey paper that provides an overview of
\emph{some} of the work that have been done in different aspects
for XML data management. Recently, the topic of large scale data
management on cloud computing and parallel data processing (e.g.
MapReduce) have been introduced and they attract a lot of interest
from the database research community~\cite{CloudDataManagment}. As
a consequence, a new series of research conferences, the ACM
Symposium on Cloud Computing, has been started in
2010~\cite{SOCC2010}. This series is co-sponsored by the ACM
Special Interest Groups on Management of Data (ACM SIGMOD) and on
Operating Systems (ACM SIGOPS). The conference will be held in
conjunction with ACM SIGMOD and ACM SOSP Conferences in alternate
years.

\section{Conclusions}
\label{SecConclusions}

Research is a competitive endeavor. Research scholars usually have
multiple goals to achieve and it is therefore reasonable that
their impact must be judged by multiple criteria. We believe that
ranking of research scholars based on the count of their
publications in top-tier research venues can be an insightful
indicator in a comprehensive assessment process. Other important
factors such as: invitations to program committees of prestigious
conferences, membership on editorial boards of high quality
journals, grant funding and awards can be also good indicators for
evaluating the impact of research scholars.

In this paper, we presented a detailed study for the publications
of 6 major database conferences and 3 major database journals in
the period between 2001 and 2010. The results of our study reveals
the fact that the number of research publications pear year and
the community size has nearly doubled through the last decade. The
results also show a considerable overlap between the top
publishers lists of the core database technology conferences and
the database journals. The results are also compatible with the
fact that the researchers in the database community tend to prefer
publishing their work in prestigious conferences rather than in
major database journals. The average publication rate for top
publishers in conference venues highly exceed their average
publication rate in the major database journals. In principle, we
believe that conference publications will remain as an attractive
way to gain a quick publicity for new research findings. However,
the number of conferences or the number of accepted publications
per conference can not continue increasing as this will limit the
value of these venues gradually. Therefore, we believe that
journal papers will remain as the best way to document and archive
significant pieces of research which can not fit within  the
12-page limit of conferences. The community should continue
pushing towards achieving the switch to the culture of highly
evaluating the journal papers over the conference
papers~\cite{Growup}. One of the valuable trials in this direction
is the introduction of the The Proceedings of the VLDB Endowment
(PVLDB)\footnote{http://www.vldb.org/pvldb/} which aims of
providing \emph{journal-like} experience to authors of the VLDB
submissions.

\bibliographystyle{plain}
\bibliography{Biblio}
\end{document}